# Design Considerations for a 5G Network Architecture

**Steven Bergren**

Oklahoma State University, Stillwater

**Abstract.** The data rates of up to 10 GB/s will characterize 5G networks telecommunications standards that are envisioned to replace the current 4G/IMT standards. The number of network-connected devices is expected to be 7 trillion by the end of this year and the traffic is expected to rise by an order of magnitude in the next 8 years. It is expected that elements of 5G will be rolled out by early 2020s to meet business and consumer demands as well as requirements of the Internet of Things. China's Ministry of Industry and Information Technology announced in September 2016 that the government-led 5G Phase-1 tests of key wireless technologies for future 5G networks were completed with satisfactory results. This paper presents an overview of the challenges facing 5G before it can be implemented to meet expected requirements of capacity, data rates, reduced latency, connectivity to massive number of devices, reduced cost and energy, and appropriate quality of experience.

## Introduction

Mobile communications have changed from a system capable of transmitting voice conversations for millions of users to a system supporting trillions of devices, primarily transporting data rather than voice. The needs for data communications are different than that of voice, and the current system of 4G/Long-Term Evolution (LTE) networks will be at maximum capacity in the next few years. New technologies and new network architectures will be required to meet the needs for future users – and there will be many of them (e.g. Andrews et al., 2014). Wireless World Research Forum predicts that there will be over 7 trillion network-connected wireless devices by the end of 2017. (Wang 2014, 123) Experts predict that traffic will be 1,000 times higher than today by 2025. (Dahlman 2014, 43) This will drive the need for higher data rates yet reduced energy consumption. These goals cannot be accomplished using the current 4G technology and architecture. Six main challenges are required to be solved in order for 5G to meet expected requirements, and many new technologies and architecture-types are believed to be able to accomplish that.

## Limitations of 4G/Challenges for 5G

The vast majority of the speed and capacity gains made in the last decade were through improvements in the efficiency of air-spectrum use (such as modulation and coding schemes) and through spectrum acquisition. (Li 2014) These methods are nearing their limit, so other methods for increase will have to be explored. Current data rates are 1 GB/s for slow or stationary devices, or 100 Mb/s for mobile devices. 5G seeks out a 10 GB/s data rate. (Wang 2014, 123) Again, technology will have to change in order to meet that demand. Base stations account for 70 percent of cellular operators' electricity usage.  Increasing capability would only increase that percentage, which is unsustainable.

4G is considered to be a descendent of 2G and 3G.  Like the evolutions between earlier generations, 4G took elements of 3G and made them better.  Instead of maintaining both circuit switching and Internet Protocol (IP) packets, 4G moved to IP for all services.  Advanced radio technologies such as orthogonal frequency-division multiplexing (OFDM) and multiple-input multiple-output (MIMO) antenna arrays were incorporated to make better use of the current spectrum. (Wang 2014, 122)  Better Quadtrature Amplitude Modulation (QAM) techniques also increased data throughput.  These advances brought data transmission speeds into the range high enough for multimedia, especially streaming video.  Now that video service is available, mobile broadband use skyrocketed.  That led to increasing smart phone usage, and many devices began to be developed.  Large format, high-resolution screens nearly always-on and always-communicating with base stations has led to high power usage with battery technologies that have met their limits.

It is clear that the many small changes made to advance through the previous generations will not be enough to make the jump to 5G requirements.  Passage from first generation through 2G, 3G, and 4G may have been evolutions on the same theme, but many of the techniques that have increased capabilities up to this point are nearing their limits.  It seems that a revolution in the mobile communications system must take place.  Drastic changes to the entire architecture of cellular networks will have to be made.

## Main Challenges

The standards for 5G have not yet been created, with future approval expected to occur in 2020.  However, telecommunications companies and the International Telecommunication Union have been working on requirements for the next generation since the last standardization of the International Mobile Telecommunications – Advanced standard in 2008, but specifically for 5G since 2015. (3GPP 2015)  Much of the problem creating the need for a new generation of mobile telecommunications comes from a wide expansion of use cases.  What once only needed to transmit voice grew to data.  And what was once just a handheld device need is growing into many disparate devices, many of which have no human interaction whatsoever.  These new use cases, which will be detailed later in this paper, have created a great number of challenges that the current system cannot support, and simple expansion of the current technologies do not seem to be a solution for.  The goals of 5G are to support these many new use cases.

### More Capacity

5G network architects predict the need to support a 1000-fold increase in traffic compared to 2010.  Mobile traffic grew nearly 70% during 2013, reaching 2.5 exabytes a month.  They expect it to grow to 10 times that monthly amount by 2019.  Theodore Rappaport of the New York University Wireless research center says 4G "can never accommodate this new demand." (Young 2015) 4G has gotten to where it is capacity-wise through radio technology advances and acquisition of new spectrum (Figure 1.) These methods are meeting their limits. Most modulation techniques and coding schemes cannot get more efficient, and the spectra best suited for long distance communications has been allocated. Though there is planned to be additional spectra available for use, the advances in capacity are projected to come from other sources. A modified expression of the Shannon theory shows that total capacity of the system is

$$C_{\text{sum}} \approx \sum_{\text{Networks}} \sum_{\text{Channels}} B_i \log_2 \left(1 + \frac{P_i}{N_p}\right)$$

where subscript $i$ is the $i^{th}$ channel, $B$ is bandwidth, $P$ is signal power, and $N$ is noise power. This shows that the total capacity can be increased via more networks (cells), channels (better spectral efficiency), bandwidth (additional spectra), and better signal-to-noise ratios. (Wang 2014, 124)

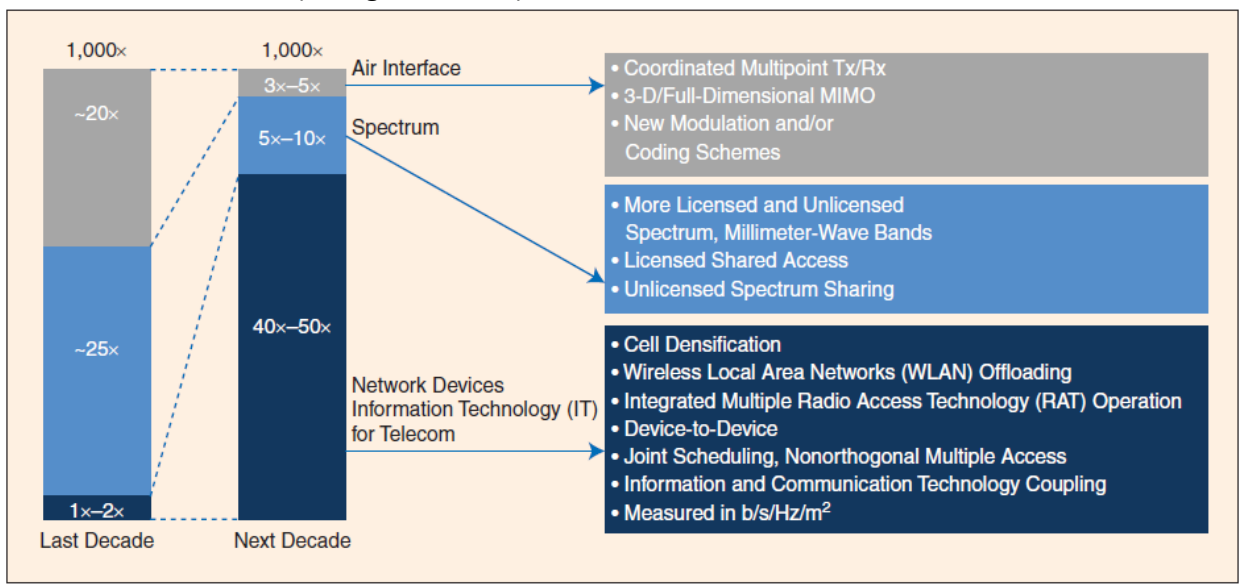

Figure 1 - Network Capacity Growth (Li 2014, 72)

### Higher Data Rate

5G goals include data rates at least ten times higher than current peak data rates. Current data rates are limited for the same reasons that capacity is limited. Increased usage of the same frequency space limits the amount of bandwidth that can be used per device. Additionally, the current capabilities of backbone, backhaul, and fronthaul limit the amount of data that can pass from base stations to mobile devices and vice versa. (Agyapong 2014, 67) Solutions will require expansion beyond current means rather than more efficient usage of the current means to affect data rate.

### Lower Latency

The Internet of Things (IoT) is putting a higher demand on always-on and always–available abilities of wireless networks. Connected devices in traffic control, autonomous cars, healthcare, and industrial automation cannot be subject to latency issues when lives may be on the line. This extremely-high-reliability and extremely-low-latency requirement is not supportable by the current model. (Agyapong 2014, 66) Previous generations of mobile telecommunications and their users primarily used static or latency-independent data, such as websites and email. Higher data rates led then to streaming video, which can be affected by latency, though buffering can prevent that. But the expansion of real-time applications, such as teleconferencing and time-critical communications, such as remote controlled medical robots, cannot suffer the effects of latency to properly function.

### Connectivity for Massive Number of Devices

Not only are the numbers of devices exploding due to IoT, but the different connection needs makes this a wide-ranging challenge. No longer are the needs expected to be similar to previous requirements (voice calls, <150 ms latency acceptable.) Devices need to be able to range from sleeping until needed to always-on, yet get the connectivity when required. (Agyapong 2014, 68) Handling the massive number of devices that may pop in and out of connection constantly will be a challenge. 4G is not ideal to handle this because current systems are disconnected, according to Zhiguo Ding of Lancaster University. (Hellemans 2015) Bluetooth, RFID, and other various short-distance communication protocols are not set up to communicate with each other. There is no common system. The challenge will be creating a system that is common yet able to support a myriad number of devices with hundreds of different use cases.

### Reduced Cost and Energy

Energy storage capabilities are not keeping up with the media abilities of current devices, and the desire to increase those abilities will require a reduction in the amount of energy required in order to communicate with the network. The IoT also is affecting this issue because small, low-power devices such as sensors are expected to run on a battery for several years. (Dahlman 2014, 42)

On the provider's side, base station electricity usage and the cost of equipment upgrades to handle the new requirements is a concern that needs to be solved for the business to remain profitable. Current estimates of the percentage of radio network power usage are 70 to 80 percent of total operations energy usage. (Agyapong 2014, 68) This is driving costs higher, and future usage will only increase that. Systems will need to change in order to meet service demands yet maintain reasonable costs.

### Quality of Experience

Quality of Experience (QoE) is the user's perception of how the experience of using a device meets expectations. It is related to some of the above challenges, such as data rate and latency, but is combined in such a way in that it creates a balance between them. High data rate and low latency may make a streaming video look good, but it is a larger drain on the battery. However, a low QoE will cause a user to become dissatisfied. Due to 4G architecture, it is difficult to maintain consistent experience at all times and in all locations.

# 5G Design Proposal Elements

Many proposals have been produced to improve the technology and architecture of the cellular network in order to achieve the goals of 5G capabilities. Though the proposals vary, many suggested elements are common across multiple papers and are possible solutions to the primary goals of 5G. These solutions, combined to work in coordination with each other, have the potential to meet the demands of mobile wireless usage for the next decade. Individual solutions, as well, are even able to meet several goals and overcome multiple challenges on their own. However, many of the proposed changes to the telecommunications system will drive additional modifications to how the network is constructed or functions in order for them to work. They need to work in concert with the other proposals in order to fully meet the expected requirements for 5G telecommunications.

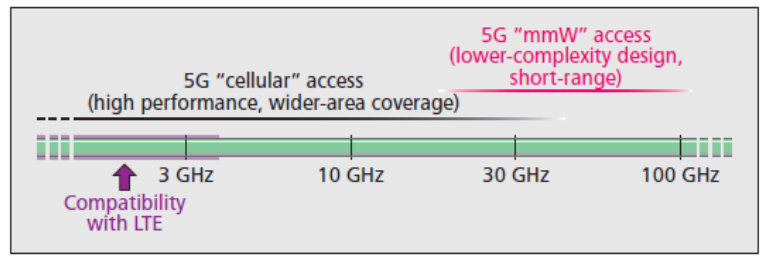

Figure 3 - Current 2G, 3G, 4G, & LTE-A spectrum and bandwidth allocation (T. Rappaport 2002)

## Spectra

Figure 2 - Spectrum range to be considered for 5G wireless access (Dahlman 2014, 45)

Much of the spectrum below a few gigahertz is already heavily populated. These frequencies are allocated, and mostly licensed, for other uses (radiolocation, satellites, radio and television broadcast, etc.) The current range for mobile telecommunications is between 700 MHz and 2.7 GHz. (Figure 2) Expansion into higher frequencies (>3GHz), and even millimeter wave frequencies (>30GHz) will be required simply because there is no more room for lower frequency allocation. (Figure 3)

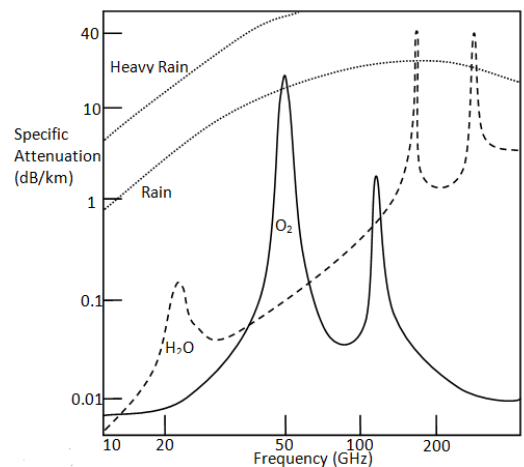

However, these higher frequencies have problems with signal propagation over large distances. There are several factors involved in this attenuation. First, signals naturally attenuate over distances travelled through over-the-air transmission. This attenuation is proportional to the square of the frequency of the signal. Therefore, entering into these higher frequency ranges will lower the range,

Figure 4 - Atmospheric Attenuation - modified from (Radiocommunications Div., U.K. Dept. Trade Industry 1988) and (T. S. Rappaport 2013)

assuming transmission power remains constant. But additional factors make these frequencies even worse. Common atmospheric gases, such as oxygen and water vapor, strongly absorb signals of specific frequencies. Rain fall is an even greater obstruction than the gases, causing nearly 10db/km attenuation at all super high frequencies and above. Most damaging to frequency propagation are solid materials. Even travelling short distances, gigahertz signals lose significant power travelling through materials such as brick, tinted windows, interior walls and cubicles. Power losses could easily be over 40 dB of attenuation. (T. S. Rappaport 2013, 340)

| Environment | Material | Thickness (cm) | Penetration Loss (dB) |
|---|---|---|---|
| Outdoor | Tinted Glass | 3.8 | 40.1 |
| | Brick | 185.4 | 28.3 |
| Indoor | Clear Glass | <1.3 | 3.9 |
| | Tinted Glass | <1.3 | 24.5 |
| | Wall | 38.1 | 6.8 |
| | 3 Walls | N/A | 41.1 |
| | 3 Walls, 2 Doors | N/A | Weak Signal Detected |
| | 5 Walls, 2 Cubicles, 1 Elevator | N/A | No signal Detected |

Table 1 – Penetration losses at 28 GHz – modified from (T. S. Rappaport 2013)

Yet another characteristic of millimeter waves that will cause problems is diffraction. Millimeter waves do not "bend" around obstructions more than a few centimeters, limiting effective usage to line-of-sight (LOS) operations. (T. Rappaport 2002, 111) Utilizing these super high frequencies and millimeter frequencies for mobile telecommunications will require other architecture changes, because the distances these signals can travel will be greatly reduced, especially inside of buildings, where 80 percent of the mobile users are. (Wang 2014, 123) These losses could be overcome by increasing power, but reducing power was a goal of 5G, so other avenues will have to be used. The solutions to these challenges brought on by super high frequency spectra usage will be discussed in later sections.

Better usage of currently-available spectrum, both licensed and unlicensed is also a tool to be used, as well. Though these frequencies are already heavily utilized, advances can still be made to increase their efficiency. First, usage of unlicensed spectrum should be considered. It can be utilized in specific cases, such as control signals, that would not be affected as much by congestion. Licensed-shared access (LSA) is another tool to be used. LSA is utilizing frequencies reserved for other purposes, as long as the usage does not interfere with the reserved purpose. These bands of frequencies often go unused for large portions of time. Proper usage will better utilize the available frequencies and increase capacity for the mobile network. Full duplex utilization of the frequency is another method to increase efficiency of already-available spectra. (Dahlman 2014, 45)

Increasing the range of spectrum used for mobile telecommunications, as well as more efficient use of currently existing frequencies, addresses both the challenges of increased capacity and increased data rate.

## Massive Multiple-Input Multiple-Output

Multiple-Input multiple-output (MIMO) antenna arrays are currently in use, but generally only have a few antennas.  MIMO arrays contain multiple antennas which can be used to beamshape and direct transmitted signals (and obtain received signals) in a specific direction.  This technology can increase the channel efficiency, increasing data rates as well as lowering energy usage as lower-power signals can be transmitted and received.  At current mobile frequencies, wavelengths are on the order of about 1 foot long.  So the half-wavelength size of the antenna is about 6 inches.  There simply isn't much room for more than a few antennas.  However, if the super high frequencies and millimeter wave frequencies proposed for 5G begin to be utilized, antennas could be on the order of a centimeter or less.  Many more antennas could be packed easily inside of handheld mobile devices than could ever fit before.

Massive MIMO (mMIMO) will have tens or hundreds of antennas and not only at the base station but dispersed throughout the cell.  These multi-antenna arrays will be placed in well-planned, line-of-sight locations to best utilize their properties.  They can be placed on buildings, connected to their internal wireless access points, which will be discussed in a later section. (Gupta, 1208)  The antenna arrays can steer beams toward a connected device, reducing power needed to communicate while also producing a low-interference signal.  The beams can also be steered into using the reflective properties of millimeter waves to circumvent the limitations of the normally line-of-sight signals. (T. e. Rappaport 2014, 111) Signals can intentionally be directed into buildings and other solid objects in order to reflect their signal into an area that the antenna cannot reach directly due to physical obstructions.  These directed signals provide a strong link that is not subject to as much signal fading as current mobile communications systems experience.  Currently, when a device is in a location of obstruction, multipath signals can then enter from the base station and act destructively, and the device must wait for a different transmission channel to send on.  This is known as fading, and it is the primary cause of latency in telecommunications.

Massive MIMO can be used to directly meet the goals of 5G by increasing the data rate and decreasing latency.  It also indirectly meets the goals by allowing the millimeter wave frequencies to be practically useful despite their characteristics of high attenuation and low diffraction.

## Device-to-Device Communication

Device-to-device communication (D2D) is the method of devices transmitting and receiving data between each other on the user plane without having to utilize network resources.  Conventional communications has all devices communicate directly

with – and only with – the base station. (Figure 5) D2D would allow multiple scenarios in which devices communicate with each other and the base station, sometimes extend cell range with an ad hoc network of hops from device to device. Devices communicating directly with each other will reduce the energy required to communicate due to shorter distances travelled and less communication due to the direct path utilized instead of relaying through the base station first. D2D can also free frequencies and base station overhead if it is not communicating with the base station for all communication, thus increasing capacity in the cell. D2D can improve data rate and lowers latency. Data can even be sent across multiple devices, with the middle devices acting as relays between the source and destination devices. (Figure 6) D2D can also reutilize spectrum, increasing efficiency. This relaying-between-devices has the added benefit of again making up for the diffraction deficiencies of millimeter waves. Devices obstructed from line-of-sight view of an antenna can have their signals relayed through a device that is in LOS with an antenna, ensuring a good connection no matter the location.

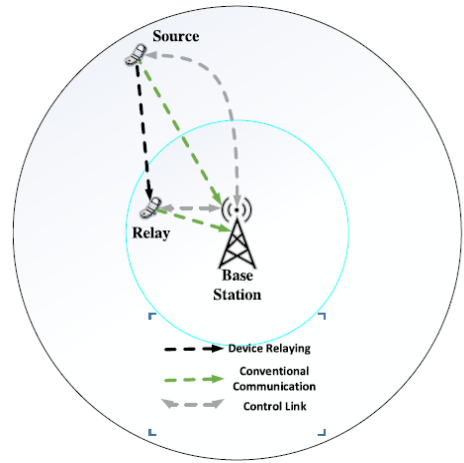

Figure 5 - Device relaying communication with base station controlled link formation. (Gupta 2015, 1218)

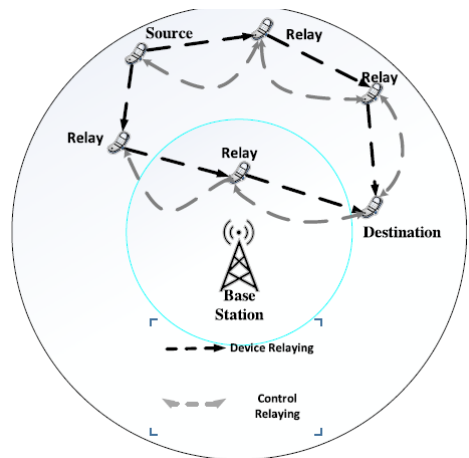

Figure 6 - Device relaying communication with device controlled link formation. (Gupta 2015, 1219)

D2D communications can contribute greatly toward meeting the goals of 5G. Lower base station usage saves energy, frees frequencies for more capacity, decreases latency, increases data rate, and like mMIMO, helps to lessen the limitations of millimeter wave LOS issues. All of these gains work together to meet user quality of experience.

## Reduced Protocol Overhead

Reducing overhead in nearly any scenario is a means to gain efficiency. Even with the simplest of mobile data transmissions, many functions take place behind the scenes that ensure that the data is sent correctly. Outside of even data transmissions, control packets are sent back and forth between base stations and devices regularly. These packets often have no payload, yet account for 46% of transmitted packets. Compressing these packets could double the efficiency of the network. (Li 2014, 75) Reducing even the amount of these packets being sent would be beneficial especially to IoT devices. (Dahlman 2014, 43) Small sensors, often running on batteries, need to conserve as much energy as possible. Constant chatter with the base station simply transmitting control packets would quickly deplete energy sources.  Being able to configure how and when devices communicate with the base station is vital for small IoT devices.

D2D communications will reduce communications between devices and base stations and data centers and all of their associated overhead will be reduced with it.  Multipath Transmission Control Protocol (TCP) can be incorporated to allow devices to utilize whatever connection is available, without having to create new paths constantly.  Simplifying protocols can also reduce the amount of data having to travel on backhaul, thereby increasing data rates.

Reducing the protocol overhead meets 5G goals by increasing data rate, reducing power consumption, and supports the IoT especially for low-power devices.

## Heterogeneous Cell Architecture

Current mobile networks are fairly homogeneous.  Though the different cells may be sized differently depending on usage density, they still communicate the same way with all devices on the network.  One of the most likely solutions for 5G, especially considering the super high frequencies that will be utilized, is to have many more cells than currently are in use and for them to be of different types and have different functions.  The millimeter frequencies proposed for 5G usage will require access points much closer to devices than current network base stations are located.  Due to the penetration issues with millimeter wave signals, network designers are considering how to handle indoor and outdoor networks separately.  One could consider the low penetration ability an advantage in that indoor systems are unlikely to experience interference from outside sources.

Current mobile cell structure is often symbolized with a honeycomb shape, showing cells that are somewhat equidistant from each other, with some small overlap to ensure connection at all points in the network.  The proposed 5G network will change that structure.  Large cells, similar to the current networks, will be called macrocells.  However, there may be many smaller cells in the range of that macrocell, as well as antennas spread throughout the cell that will communicate with the base stations via fiber optics.  Access points called femtocells will need to be put inside of buildings to avoid the signal loss of the high-frequency signals unable to get through outer walls. (Figure 7)  They can also be used in trains, buses, and cars.  This reduces the frequency usage because all devices connected to the femtocell will be seen as one unit to the base station. (Wang 2014, 124)  This also requires the networks to be able to communicate with many types of systems.  Femtocells may use WiFi or other technologies for these short-range communications, utilizing unlicensed frequencies.  These systems would then have to communicate with the base station, so the network's ability to communicate across different platforms will be essential.

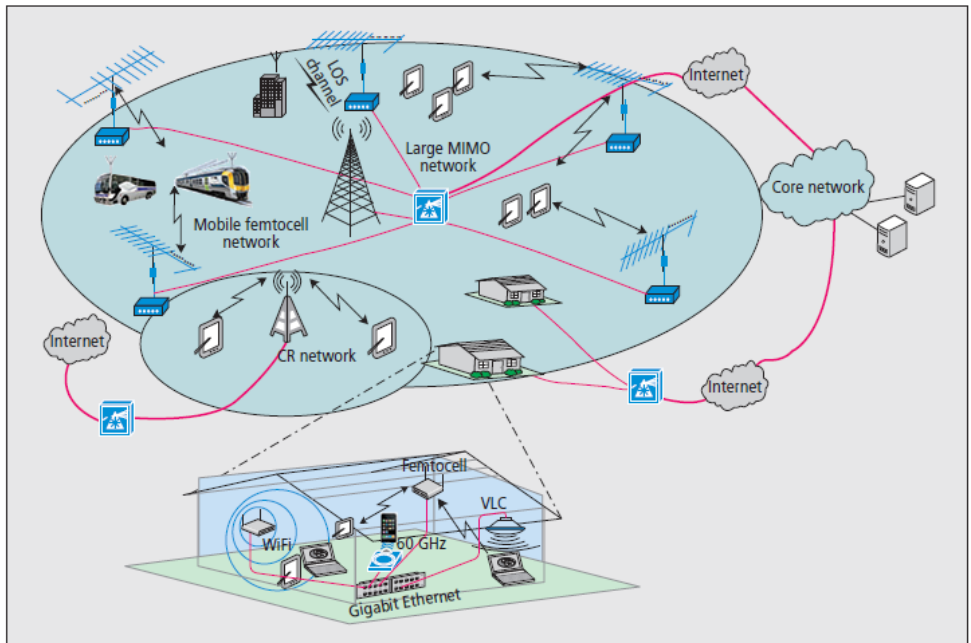

Figure 7 - A proposed 5G heterogeneous wireless cellular architecture (Wang 2014, 125)

Other ways to improve communication include dividing up the control and user data planes.  Control can be handled at the macrocell level and user data at the smaller cell.  The femtocell will taking on the majority of the data transfer, leaving the macro cell with more available resources.  By leaving the control functions at the macro cell, the femtocell is free to work primarily on data transmission, which is the primary need of the user.

Like frequency, power is related to the square of the distance.  As these cells become smaller, the distance from access point to device is reduced, further reducing the power required to transmit.  Shorter distances also decrease the chance of obstructions interfering with signal between devices and towers.  The number of antennas in the MIMO array at small cell access points also has a large effect on the amount of power used when communicating with the base stations. (Figure 8)

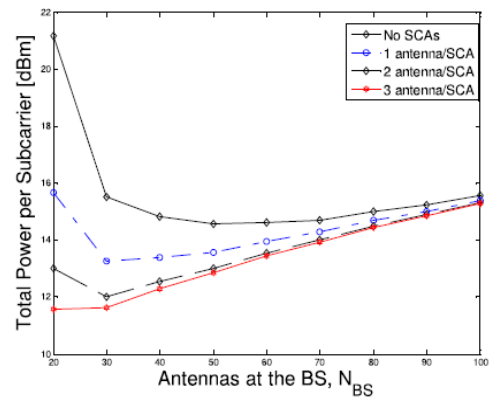

Figure 8 - Average total power consumption in the scenario containing small cell access points. (Gupta 2015, 1216)

A heterogeneous network, one of the main concepts in proposed 5G architecture, contributes to the goals of 5G by increasing capacity, increasing data rates, and reducing power consumption.

## Network Function Virtualization

A software-defined network solution is also proposed to increase efficiency as well as reduce costs.  Currently, many of the functions of control and network management are located at the base stations.  Upgrades or changes to functions are difficult and costly to implement.  By removing specialized hardware at the base stations and virtualizing them in the cloud (Figure 9), upgrades can be made more frequently and at lower cost.  Management of the entire network can be handled more efficiently at a data center, allowing downline base stations to handle only what is required at their end.  Analysis at the data center level can also include network intelligence to better route data and manage connections.

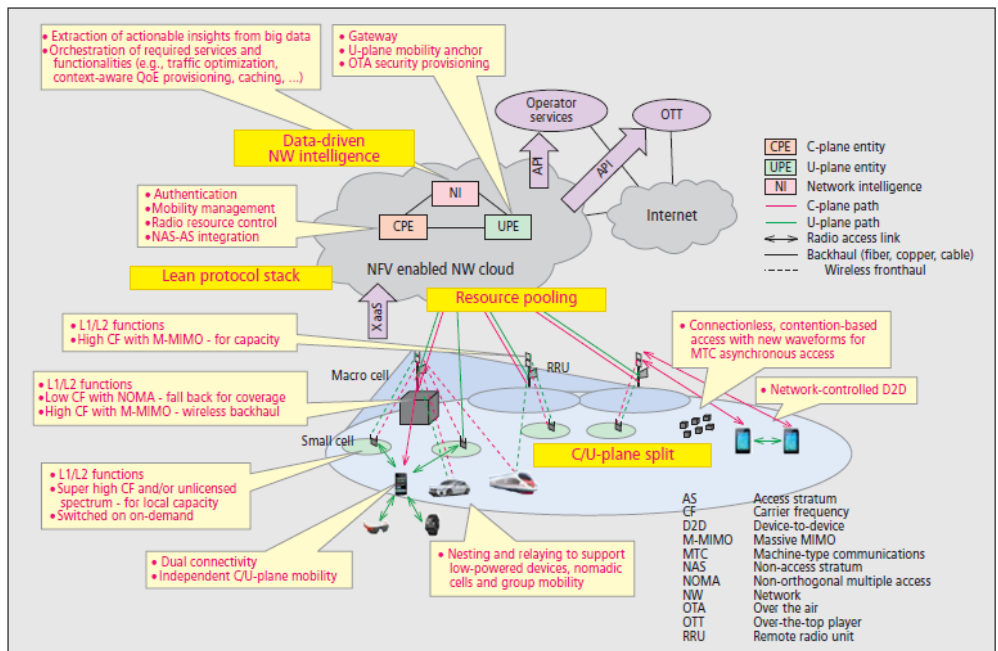

Figure 9 – Network Architecture highlighting cloud-based management (Agyapong 2014, 72)

Passing devices from base station to base station as they move between cells would be much simpler.  Currently, base stations take care of that between each other, which updated their own databases and other central office databases.  If controlled at the core network, it could free base stations of that responsibility and lessen the overhead involved. (Yazıcı 2014, 77) Deep Packet Inspection (DPI) is a tool used to see the type of data being transmitted.  By knowing the contents of packets, the network can better route the data by altering connections so that the user's quality of experience can be increased. (Yazıcı 2014, 78) This sort of function needs to be done at a data center.  It is too late in the flow of data for it to be done at the base station, so the quality of a streaming video, for instance, cannot be increased at the base station level.

Network Function Virtualization supports the 5G goals by decreasing costs and increasing user quality of experience.

## Content Caching

More intelligence at the edge of the network, however, can also be helpful.  Just as Deep Packet Inspection can increase quality of experience at the core network level, content caching at the base station level can likewise add to that quality.  Pre-fetching or caching data at lower levels can reduce backhaul usage, lower latency and increase the QoE of the users. (Agyapong 2014, 68) For instance, a base station may call for data for a commonly requested, latency critical item be downloaded and cached at the base station so that when a user requests the data, it does not have to travel through the backhaul system and the rest of the internet, which being a "best effort" service, will not necessarily meet the expectations of the user.

Content caching can increase quality of experience for users by reducing latency.  It also increases apparent data rate by moving the data closer to the user and off congested backhaul paths.

## Practical Considerations and Infrastructure Requirements

The many technological advances addressed in the previous section are viable solutions to the problems and challenges stated prior and may be able to meet the proposed requirements for 5G networks when considered all together.  However, turning these possible solutions into actual, implemented hardware and infrastructure will be a gargantuan task with many obstacles in the way.  These types of concerns are rarely discussed in scholarly articles; however they are a critical concern in the deployment of mobile networks, being a consumer business, and factors greatly into the design of the system.  In a Global mobile Suppliers Association survey of the industry on the challenges to meeting the 5G goals, most all surveyed scored the list of barriers at least at a 3, and most close to a 4.  Cost, lack of suitable spectrum, and delay of standards/interoperability were the three highest ranked issues. (GSA 2015, 29)

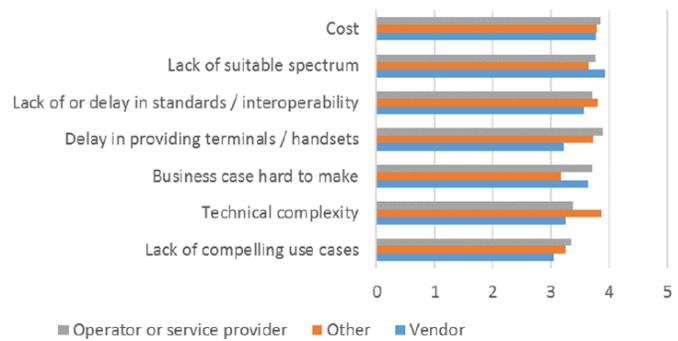

Figure 10 – Main barriers to 5G development and deployment (5 = very serious barrier; 1 = not a barrier at all) (n=97)

### Cost

Previous generations of mobile technology certainly had increased costs when transitioning, but for the most part, the changes were minimal.  Hardware needed to be changed at the base stations and central office, but little more was done.  The solutions considered in this paper will require an enormous increase in infrastructure costs.  Network virtualization of functions will require large upgrades of the core servers to control the network.  Changes will need to be made at every base station in order to accommodate the cloud-based computing and control of the towers.  Perhaps most costly will be the additional small cell deployments.  According to wirelessestimator.com, there are currently about 120,000 cell towers in the US. Assuming only one additional small cell per current tower, that doubles the current number of towers.  These new cells may not need full size towers, however, because they may be placed on buildings and will be located in more densely populated areas.  But that does not negate the fact that the available real estate must be found, then purchased or leased, properly zoned, and power and backhaul utilities run to the locations.  (DeGrasse 2015)

A look back at Figure 7 (page 9) and Figure 9 (page 11) shows how the current paradigm (the one large macro cell tower) is augmented with multiple antennas throughout the macrocell and femtocells on trains, buses, and buildings.  There is also a reliance on femtocells in homes routing their wireless communications through hardwired internet connections.  To get the speeds of 5G, houses will have to have fiber run to every home, which is usually only done today in brand new developments because ISPs do not find it cost effective to run it everywhere.  Most homes have only copper connections.  The goals of ultra-low latency may be great, but that is only on the radio network side.  The overall latency of the user is a factor of the radio network, but also of all other connections on the internet to get the content desired.  The cost to get low latency on the radio network, which is only part of the entire latency string, is not something most consumers would want to pay for. (Lind 2016)

## Lack of Suitable Spectrum

Government entities that regulate the radio spectrum in their countries have control of what and how much of a frequency can be allocated to mobile telecommunications.  The process to open new spectra can sometimes be a slow, arduous task that does not keep up with the speed at which industry desires.  It may also be done in a way that hampers the full and fair usage of that spectrum.  In the US, the FCC recently opened several large sections of high frequency spectrum for 5G usage.  However, some fear that it is not enough and that the big carriers will license portions that they will then not use completely, only covering large city centers and venues with 5G and leaving the rest of the country without. (Goovaerts 2016) This locking down of frequencies will prevent true and full deployment of 5G in all areas.

## Delay of Standards / Interoperability

The standards for any technology define what can and can't be done with it.  These standards then define how carriers will design and implement their networks.  That design then leads to hardware designs (both the radio network and mobile devices) using the standards that are then manufactured, and then ultimately deployed.  Though some capital can be expended prior, it is not necessarily in the best interest of the carriers to go out early on their own and presume certain 5G requirements.  The more complicated the 5G design, the more important it is to get a solid standard, and that takes time. (Kinney 2015) The amount of interoperability and backwards compatibility will also be a concern.  If these networks have to support 4G, 3G, and possible prior generations, it will affect how the system is designed to be able to still communicate with the older technologies, yet remain able to meet the goals of 5G.

The reliance on other communication types, such as WiFi, device-to-device communication and consumer internet service provider access, will cause an entire other set of concerns.  Those networks will be required to work together in order to meet the 5G goals.  The entire industry of communications, from wireless carriers to wired carriers, and the manufactures of all the devices used by those carriers, will have to come together and agree on how to operate interchangeably across the different networks and establish rules for cost accounting and sharing.  These agreements will only add on to the delay created by a late standard.

## Conclusion

Many new technologies and a redesign of the wireless network architectures will be able to make the goals of 5G realizable in the near future.   The ability and desire for mobile telecommunication companies to implement all of the solutions, however, may require additional solutions to be found (e.g. Bjornson et al., 2015). For example, millimeter wave communication technologies may be employed for backhaul traffic (Ge et al. 2014). Another idea is to decouple downlink and uplinks (Elshaer et al., 2014). Location information can improve scalability, latency and robustness of 5G (Taranto et al., 2014). There are also issues related to security given the massive number of connections and quantum techniques (Kak 1999; Kak 2007) might be useful in certain situations especially where one wants to simulate new primitives such as entanglement (Kak 2016).

 There may also need to be a balance created between these various solutions, depending on the physical environment and capacity density desired.  Due to these considerations, portions of 5G requirements may be implemented first, such as additional small cells, before other parts, such as network virtualization, are implemented.  A true, fully-implemented-to-the-standard 5G may not be realized for quite some time, and even then, perhaps only in areas that carriers believe they will get the most out of their deployment costs.

Whatever the actual implementation, the technologies are being created and the capability is being developed to be able to implement the 5G requirements in order to meet the mobile communications demands of 2020 and beyond.  Early successes in test deployments will further enable modification of systems so that a functional system will be available in the next several years.